\documentclass[10pt,letterpaper]{article}
\usepackage[top=0.85in,left=2.75in,footskip=0.75in,marginparwidth=2in]{geometry}

\usepackage[utf8]{inputenc}

\usepackage{cite}

\usepackage{nameref,hyperref}

\usepackage{microtype}
\DisableLigatures[f]{encoding = *, family = * }

\raggedright
\usepackage{changepage}

\usepackage[aboveskip=1pt,labelfont=bf,labelsep=period,singlelinecheck=off]{caption}

\makeatletter
\renewcommand{\@biblabel}[1]{\quad#1.}
\makeatother

\usepackage{lastpage,fancyhdr,graphicx}
\usepackage{epstopdf}
\fancyheadoffset[L]{2.25in}
\fancyfootoffset[L]{2.25in}

\usepackage{mathrsfs,amsmath}
\usepackage{amssymb}

\usepackage{float}

\usepackage{color}

\definecolor{Gray}{gray}{.25}

\usepackage{graphicx}

\usepackage{sidecap}

\usepackage{wrapfig}
\usepackage[pscoord]{eso-pic}
\usepackage[fulladjust]{marginnote}
\reversemarginpar

\begin{document}
\vspace*{0.35in}

% title goes here:
\begin{flushleft}
{\Large
\textbf\newline{Detection of Multidecadal Changes in Vegetation Dynamics and Association with Intra-annual Climate Variability in the Columbia River Basin}
}
\newline
% authors go here:
\\
Andrew B Whetten \textsuperscript{1}, Hannah Demler \textsuperscript{2}
%Author 2\textsuperscript{2},
\\
\bigskip
\bf{1} Department of Mathematical Sciences, University of Wisconsin- Milwaukee, WI
\\
\bf{2} USDA ARS Ainsworth Research Lab, University of Illinois - Urbana Champaign, IL
\\
\bigskip
* awhetten@uwm.edu

\end{flushleft}

\section*{Abstract}
Leaf Area index is widely used metric for the assessment of vegetation dynamics and can be used to assess the impact of regional/local climate conditions. The underlying continuity of high resolution spatio-temporal phenological processes in the presence of extensive missing values poses a number of challenges in the detection of changes at a local and regional level. The feasibility of functional data analysis methods were evaluated to improve the exploration of such data.  In this paper, an investigation of multidecadal variation of leaf area index (LAI) is conducted in the Columbia Watershed, as detected by NOAA AVHRR satellite imaging, and its inter- and intra-annual correlation with maximum temperature and precipitation using the ERA-Interim Reanalysis from 1996 to 2017. A functional cluster analysis model was implemented to identify regions in the Columbia Watershed that exhibit similar long-term greening trends. Across these several regions, the primary source of annual LAI variation is a trend toward seasonally earlier and higher recordings of regional average maximum LAI. Further exploratory analysis reveals that although strongly correlated to LAI, maximum temperature and precipitation do not exhibit clear longitudinal trends.

% now start line numbers
%\linenumbers

% the * after section prevents numbering
\section{Introduction}
% broad intro- climate change
Understanding how changing climate conditions interact with and affect the biosphere is necessary for characterization of global change and for developing and implementing adaptive and sustainable practices worldwide. The study of ecological responses to climate change has greatly expanded in the recent past with the development and use of remote sensing and the compilation of multidecadal satellite data sets. Satellites allow for near continuous observations of earth and monitoring of changes in the biosphere on scales ranging from local to global ~\cite{Ustin and Middleton 2021}. 
% plant responses to global change and LAI measurements

Plant life serves a critical role globally in regulating energy, chemical, and mass transfers within the earth system and changes in vegetation at different scales effects land surface - climate system feedbacks and ecosystem services through regulation of carbon, water, and energy exchanges between the atmosphere and biosphere ~\cite{Wu 2016, Zeng 2017}. Many studies have reported a "greening" of the planet attributed to the "fertilization effect" of higher atmospheric [CO\textsubscript{2}], nitrogen deposition, or increasing temperatures lengthening the growing season in many regions ~\cite{Zhu 2016, Fensholt 2012, Mao 2016}. However, complex interactions exist between plant physiology, atmospheric conditions, temperature, and water and nutrient availability that make the picture of how plants respond to changing climate conditions less clear ~\cite{Tubiello 2007, Leakey 2014}.

Phenology refers to periodic and seasonal reproductive events in biological life cycles. Vegetative phenological phenomena are sensitive to annual climate conditions and therefore changes in phenology, such as the timing, rate, duration, and magnitude of annual vegetative growth, can signal important effects of climate change on plants ~\cite{Piao 2019}. Leaf Area Index (LAI), a widely utilized measure of plant growth and activity, is a unit-less measurement of leaf area (m\textsuperscript{2} ) per ground area (m\textsuperscript{-2}). LAI provides a key measure of plant cover in a given area and is defined as an essential climate variable (ECV) by the Global Climate Observing System (GCOS) due to its critical contribution to the characterization of Earth's climate ~\cite{Bojinski 2014}. Satellite-derived LAI products offer multidecadal records of terrestrial plant cover around the world, allowing for analysis of inter-annual variability in vegetation dynamics which provides key insight to how plants respond to global change.

% Andrew's notes so far
Earth-observing satellites have used greening indices to infer phenological changes and vegetation stress since the early 1980’s and in recent decades the quality of these products has continued to improve. These improvements have yielded massive reservoirs of high resolution spatio-temporal data~\cite{Schnase2016}.  Remotely sensed vegetation data poses a number of challenges. (1) The regular and dense occurrence of missing values resulting from excessive cloud cover, snow cover, and barren landscapes requires extensive pre-processing of the data. (2) For the analysis of large regions,  tens of thousands of sites are available. The analysis of thousands of adjacent sites requires modeling of the inherently  complex spatio-temporal correlatory structure. (3) The collection of a phenological process over multiple decades yields many replications of a stochastic process at each site. The variability of intra-site replications of an annual process is often characterized by complex cyclic or at least year-dependent structure.

This analysis is motivated by the recent work to detect summer NDVI greening patterns in the Arctic Tundra Biome conducted by the GEODE Lab at Northern Arizona University~\cite{Berner2020}. Their work utilizes maximum recorded greenness at a given site annually, as measured by NDVI, to detect greening or browning trends across 50,000 sites in the Arctic tundra. To the best of our knowledge, most literature on the analysis of greening trends rely heavily on the annual maximum or mean greenness trends using time series and temporal site/region correlation~\cite{Sumida2018, Forzieri2018, Piao2003}.

% https://agupubs.onlinelibrary.wiley.com/doi/10.1029/2002JD002848
% https://www.sciencedirect.com/science/article/abs/pii/S0303243400850114
% https://www.nature.com/articles/s41598-018-31672-3
% https://agupubs.onlinelibrary.wiley.com/doi/full/10.1002/2018MS001284
% https://journals.ametsoc.org/view/journals/clim/20/7/jcli4067.1.xml

The variations of this approach have a number of advantages. Maximum/mean greenness is by far the most important phenological attribute, and the dimensionality of the data is decreased exponentially as measurements throughout the remainder of the years are filtered out. Further, the expansive number of missing values inherently present in such data are removed. These are all advantageous and justifiable simplifications that ameliorate subsequent statistical analysis.

% THE CASE FOR FDA Methods
This simplification neglects modeling the elegant periodicity and continuity of plant phenology. Further, it limits the potential of an effective exploratory analysis of such data products. In this project, a philosophically different approach is taken to analyze remotely sensed phenological data in which the unit of analysis is a curve (or function) as opposed to a single site measurement. This approach, widely referred to as functional data analysis (FDA) roots in the assumption that measurements vary over some continuum such as space or time and that there is an underlying smoothness inherent to the process of interest~\cite{Ramsay2005, Ramsay1991, Finch2013}. A temporal process, measured discretely on a regular or irregular time grid, is smoothed using an optimized basis function expansion. In almost all circumstances, replications of this process are present at the same spatial location or across several different locations, and as such, a collection of continuous differentiable curves are obtained for analysis. 
%Refer to Fig... for an example. % probably don't need to reference figures in the Intro section

The retention of an entire smooth curve yields several advantages  for vegetation dynamics applications: (1) Missing values can be effectively smoothed over, (2) Anomaly values can be filtered/smoothed out of the data, (3) the timing of the magnitude of max greenness is retained naturally/simultaneously, and (4) differential information is naturally contained in the basis function expansion used to smooth the curves. Literature documenting the use of FDA in remotely sensed plant phenology is sparse, but recent work in mapping forest plant associations using FDA combined with other machine learning methodology has shown promising results~\cite{Pesaresi2020}.

%Columbia River Basin
The Columbia River Basin (CRB) is located in the north-western United States and south-western British Columbia, Canada.The drainage basin is bounded by the Rocky Mountains to the east and the Cascade and Coast ranges to the west and covers and area of 670,000 km\textsuperscript{2}: 568,000 km\textsuperscript{2} of which are spread across the US states of Washington, Oregon, Idaho, Montana, Wyoming, Utah, and Nevada. Climate in the CRB varies from humid and maritime along the western parts of the basin to semi-arid and arid in the southeast. The CRB hosts a range of diverse natural ecosystems as well as large agricultural regions consisting largely of forestry, dairy and cattle farming, and production of apples, potatoes, wheat, and other small grains. (USGS River Basins of the United States: the Columbia report). 
%add this citation to bib

% Brief summary of methods proposed
In this project, a cluster analysis is performed on 27,196 sites in the CRB that incorporate pairwise site correlations between smoothed multidecadal LAI site profiles from 1996-2017 remotely sensed using NOAA AVHRR times series product. Intuitive clusters were detected that are largely distinguished by land cover, elevation, the seasonal timing and magnitude of peak LAI. Further, substantial regional inter-annual variation is identified across several clusters that is explained by earlier and increasing magnitude of maximum LAI. We supplement this work with an exploratory applet that can be accessed from Supporting File S2.

Using an ERA-Interim data product, strong correlations were detected between intra-annual temperature profiles characterized by warmer temperatures during the first 20 weeks of the year and the timing and magnitude of LAI throughout the CRB region. However, Inter-annual trends in temperature over the studied time period do not match the clear long term changes detected in LAI. Variation in precipitation profiles was not uniformly correlated with timing and magnitude of LAI in the CRB region. 

%Since the AVHRR satellite have known limitations in relating observed trends to field measurement resolution, further exploration of the relationship between an array of Fluxnet site US Me2 attributes to local AVHRR LAI sites is done to improve the validation of these results and search for other driving factors of the substantial changes in LAI detected in the CRB.

These results provide an innovative framework for future analysis of remotely sensed data products using functional data and spline smoothing methods, and provides further confirmation of greening trends observed across the world in recent decades.

% Brief summary of findings

%The introduction should briefly place the study in a broad context and highlight why it is important. It should define the purpose of the work and its significance. The current state of the research field should be reviewed carefully and key publications cited. Please highlight controversial and diverging hypotheses when necessary. Finally, briefly mention the main aim of the work and highlight the principal conclusions. As far as possible, please keep the introduction comprehensible to scientists outside your particular field of research. Citing a journal paper \cite{ref-journal}. Now citing a book reference \cite{ref-book1,ref-book2} or other reference types \cite{ref-unpublish,ref-communication,ref-proceeding}. Please use the command \citep{ref-thesis,ref-url} for the following MDPI journals, which use author--date citation: Administrative Sciences, Arts, Econometrics, Economies, Genealogy, Histories, Humanities, IJFS, Journal of Intelligence, Journalism and Media, JRFM, Languages, Laws, Religions, Risks, Social Sciences.
 
%%%%%%%%%%%%%%%%%%%%%%%%%%%%%%%%%%%%%%%%%%
\section{Materials and Methods}

\subsection{LAI AVHRR Climate Data Record}

The LAI Climate Data Record (LAI CDR) produces a daily product on a 0.05 degree x 0.05 degree grid dating back to 1981 derived from Advanced Very High Resolution Radiometer (AVHRR) sensors using data from eight NOAA polar orbiting satellites: NOAA -7, -9, -11, -14, -16, -17, -18 and -19. The highest resolution of AVHRR sites is approximately 1km per pixel~\cite{Claverie2016,Claverie2014}. In this analysis, we subset the data from January 1st, 1996 until December 31st, 2017 and the spatial domain is restricted to 37,110 sites in the US portion of the CRB (refer to Fig~\ref{fig2}). In this 22-year period, daily LAI measurements are summarized on a weekly resolution, by taking weekly average LAI across a 7 day period. The resulting data product has 1152 weeks. In this product, there are thousands of sites that report high volumes of missing values. In order to construct spline smoothed curves on the 22 year period,  a minimum threshold was set for 28 percent of weeks in the 22 year period to have at least one weekly recording of LAI. This filtering process leaves 27196 sites. By inspection, it is clear than many sites of the removed sites are barren/sparsely vegetated regions and high altitude sites, but we are not certain of the quality of the products in the regions where high densities of sites report excessive missing values. From these results, it appears that this approach is robust enough to handle sites with higher occurrences of missing values (towards a threshold of 15 to 20 percent), although this is left to future work.

% Claverie, M., Matthews, J., Vermote, E., Justice, C.,: A 30+ Year AVHRR LAI and FAPAR Climate Data Record: Algorithm Description and Validation. Remote Sensing, Vol 8, Issue 3: 263, 2016.

%Claverie, Martin; Vermote, Eric; NOAA CDR Program. (2014): NOAA Climate Data Record (CDR) of Leaf Area Index (LAI) and Fraction of Absorbed Photosynthetically Active Radiation (FAPAR), Version 4. [indicate subset used]. NOAA National Centers for Environmental Information. https://doi.org/10.7289/V5M043BX. Accessed [date].

\subsection{ERA-Interim Reanalysis}

The ERA-Interim is a reanalysis of the global climate attributes covering the data-rich period since 1979 (originally, ERA-Interim ran from 1989, but the 10 year extension for 1979-1988 was produced in 2011), and continuing in real time until its discontinuation in 2019~\cite{Dee2011}. The spatial resolution of the data set is approximately 80 km and provides daily recordings of maximum temperature, minimum temperature, and precipitation. [The product used in this analysis has been pre-processed into longitudinal attributes of these climate attributes by Jupiter Intelligence (for the ENVR 2021 Data Challenge)]. This product is subsetted temporally to the same domain prescribed for the LAI CDR. The spatial domain for this product is restricted to a square region containing the subsetted points from the LAI CDR with the following boundaries ($Lat_{min}=40.5N , Lat_{max}= 49.0N, Lon_{min}=-108.0W , Lon_{max}=-124.0W$). Requiring that the subsetted LAI CDR  product fully contained within the subsetted ERA-Interim product is optimal for appropriate spatial prediction of climate attributes of the ERA-Interim product onto the LAI CDR coordinate system.

%Dee, D. P., and Coauthors, 2011: The ERA‐Interim reanalysis: configuration and performance of the data assimilation system. Q.J.R. Meteor. Soc., 137, 553-597,  https://doi.org/10.1002/qj.828.

\subsection{BaseVue 2013 Land Cover and USGS National Elevation Products}

The results of the cluster analysis prompted further investigation into site characteristics that may be driving factors in the separation of clusters. We use the BaseVue 2013 Land Cover product, which is a commercial global, land use/land cover product developed by MDA~\cite{MacDonald2014, NatElev2002}.  BaseVue is independently derived from roughly 9,200 Landsat 8 images and has a spatial resolution of 30m. The capture dates for the Landsat 8 imagery range from April 11, 2013 to June 29, 2014, and contains 16 classes of land use/land cover. Elevations were extracted at each site using the USGS National Elevation product. This dynamic image service provides numeric values on a 30m resolution representing orthometric ground surface heights (sea level = 0) which are based on a digital terrain model (DTM). 

% Mention weakness of this in the Discussion section

%Slope, aspect, and hillshade are also extracted from this project for exploratory purposes, and the details of this analysis are left to the appendix.

%MacDonald, Dettwiler and Associates Ltd. (MDA). 2014. BaseVue 2013. Available at: http://www.arcgis.com/home/item.html?id=1770449f11df418db482a14df4ac26eb [Last accessed 23 March 2021].

% National Elevation Dataset; 2002; Web site; U.S Geological Survey

\subsection{Smoothing LAI}

For a collection of raw weekly average recording of LAI, denoted by $Y = [\overrightarrow{y}_1 \ldots \overrightarrow{y}_n]$, we estimate $\hat{x}(t) =\sum_{k=1}^K c_k \phi_k(t)$ subject to a roughness penalty on the second derivative of the basis expansion $\Phi = [\phi_1(t) \ldots \phi_K(t)]$ where $c_k$ are the coefficients of the terms of the basis expansion denoted by $\phi_k$. This can be expressed as an unconstrained minimization defined by 

\begin{eqnarray}
\label{eq:schemeP1}
	\underset{\overrightarrow{c}}{min} \|\overrightarrow{y} - \Phi \overrightarrow{c}\|^2 + \lambda c^T R c \;\;\; for \;\; \lambda \geq 0,
\end{eqnarray}

\noindent where $R_{jk}=\sum^M_{l=1} \phi_j^{''}(\overset{\sim}{t}_l) \phi_k^{''}(\overset{\sim}{t}_l)h $  for $h=\overset{\sim}{t}_l -\overset{\sim}{t}_{l-1} $~\cite{Ramsay2005}. 

\begin{figure}[H]	
%\widefigure
\centerline{\includegraphics[width=0.8\textwidth]{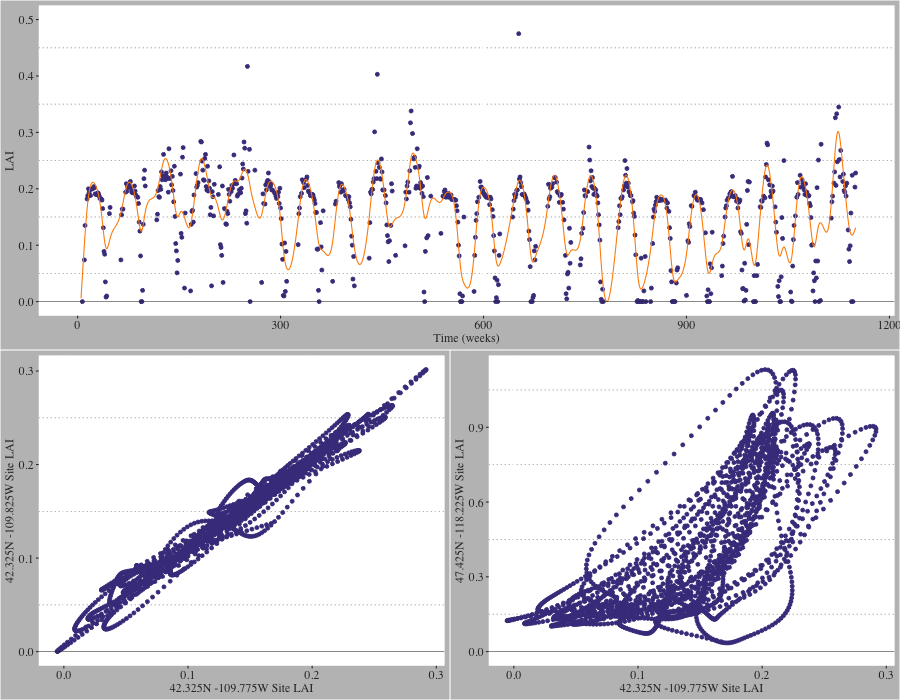}}
\caption{Illustration of the LAI smoothing process. (Upper) Raw and B-spline smoothed LAI splines overlayed for Site X (42.325N -109.775W): The spline model retains the functional structure of the raw LAI recording while filtering out anomaly/false recordings. (Lower Left) Spline smoothed LAI for adjacent sites: Site X and Site Y (42.325N -109.825W). The spearman correlation is 0.988. Site X Elevation = 2133.8m and Site Y Elevation = 2168.8m, and both sites are classified as Scrub/Shrub locations. (Lower Right) Spline smoothed LAI for Site X  and Site Z (47.425N -118.225W). The spearman correlation is 0.769. Site Z Elevation = 690.5m and is classified as an agriculture location~\cite{Wickham2016,Baptiste2015}.
\label{fig1}}
\end{figure}

We select an appropriate value for $\lambda$ coordinates using the optimal lambda for a single site determined by the generalized cross-validation criteria, $GCV = \frac{MSE(\lambda)}{(1-\frac{df_\lambda}{M})}$. This is justifiable since the roughness of LAI recording at all sites are similar, and the roughness penalty restriction permits the estimated functions to characterize the same types of features across coordinates in the Columbia Watershed. The resulting smoothed LAI curves have the form 

\begin{eqnarray}
\label{eq:schemeP2}
	\hat{x} = \Phi(\Phi^T \Phi + \lambda R)^{-1} \Phi^T \overrightarrow{y} = S \overrightarrow{y}.
\end{eqnarray}
In the upper plot of Fig~\ref{fig1}, we present an example of a smoothed 22-year LAI profile. We emphasize that the curves retain the timing and structure of the raw data while naturally filtering out singleton anomaly values that yield false maximum LAI readings.

\subsection{Spatial Clustering of High Dimensional Functional Data} \label{cluster}

Consider each smoothed curve, $\hat{x_j}(t)$ as a vector of measurements. The pairwise correlation of $\hat{x}_j(t)$ with $\hat{x}_{j'}(t)$ can be measured by plotting the two curves against each in $\mathbb{R}^2$. Plotting smoothed LAI curves in this way yields ellipsoidal paths of time-paired LAI recordings $(\hat{x}_j(t_i),\hat{x}_{j'}(t_i))$. We measure the strength of monotonicity of the ellipsoidal relationship of coordinates rather than its linearity. To accomplish this, we compute Spearman rank correlations defined by

\begin{eqnarray}
\label{eq:schemeP3}
	\rho(\hat{x}_j, \hat{x}_{j'}) = \rho_{jj'}= \frac{S_{\hat{x}_j\hat{x}_{j'}}}{S_{\hat{x}_j}S_{\hat{x}_{j'}}} =
	\frac{n^{-1}\sum^n [R(x_j) - \overline{R(x_j)}] [R(x_{j'}) - \overline{R(x_{j'})}]}{\sqrt{n^{-1}\sum^n [R(x_j) - \overline{R(x_j)}]^2 \cdot n^{-1}\sum^n [R(x_{j'}) - \overline{R(x_{j'})}]^2 }}
\end{eqnarray}

\noindent between all pairs of coordinates where $R(x_j)$ is the ranking function of the elements of  $\hat x_j(t)$. This approach proves advantageous in this application since Spearman correlation is a computationally efficient measure of association for a high dimensional quantity of paired coordinates. This approach also provides a notion of standardization of the measure of association which is of great application-based importance, namely that coordinates grouped in the same cluster have increasingly synchronous greening patterns independent of the differences in the magnitude of LAI at each site. (As an example, coordinates on north facing slopes may experience lower and seasonally delayed LAI than south facing slopes, but if close enough in proximity the regularity of paired ellipsoidal LAI movement in $\mathbb{R}^2$ will likely be high.

The $\rho_{jj'}$ can be thought of as a discretization of a spatial correlation function $P$ where $\rho_{jj'} = P(x_j, x_{j'})$. The availability of longitudinal recordings of LAI at each coordinate allows for direct estimation of $\rho_{jj'}$ without placing assumptions on $P$~\cite{Gervini2019}. The computed $nxn$ matrix of pairwise correlations is used to construct a dissimiliarity matrix with elements $d_{jj'} = 1 - \rho_{jj'}$.

By construction, pairs of coordinates with high spearman correlation have low $d_{jj'}$ values, and as such, they are considered to be close in ``distance" to each other. We then perform k-mediod clustering using the partitioning around mediods algorithm (PAM) for $k=2, \ldots, 6$ clusters~\cite{Cluster2021}.

\subsection{Ordinary Kriging of the ERA-Interim}

Since the ERA-Interim reanalysis product is on an 80km resolution, spatio-temporal prediction onto the LAI CDR grid is required. This is accomplished using Ordinary Functional Kriging on the ERA-Interim. Considering a functional random process $\{X_S : s\in D \subset \mathbb{R}^d\}$ with $d=2$ such that $X_S$ is a functional random variable for any $s \in D$ observed at $n$ sites~\cite{Giraldo2020, Giraldo2012}. It is  assumed that the random process is second order stationarity and isotropic. Attempting to predict a complete smooth function $\hat{X}_{S0}:[a,b] \to \mathbb{R}$, expressed by 

\begin{eqnarray}
\label{eq:schemeP4}
	\hat{X}_{S0} = \sum^{n}_{i=1} \lambda_i X_{S_i}, \;\; \lambda_1, \ldots, \lambda_n \in \mathbb{R}.
\end{eqnarray}

By construction, $\hat{X}_{S0}$ is a linear combination of observed curves with weights $\lambda_1, \ldots, \lambda_n \in \mathbb{R}$. Curves from locations closer to the prediction site are constructed to have increased influence on the prediction. We use a parametric estimated trace-semiovariongram with the exponential distribution to obtain the kriging weights $\lambda_i$.

\subsection{Interannual Regional LAI and Climate Variation Monitoring}\label{fpca}

Having obtained smoothed weekly average LAI, average maximum temperature, and precipitation curves from 1996 to 2017 on the 0.05x0.05 degree grid, the 22 year profiles were averaged by cluster to obtain regional average curves for the respective variables. Subsequently, the 22 year profiles were decomposed into a collection of 22 annual curves. The deconstruction into annual curves provides the needed replications of curves to assess interannual variation. For a given collection of average annual profile for a given cluster LAI, denoted by $\hat{\bar{X_i}}(t)$, and the Karhumen-Loeve decomposition of the annual profiles can be expressed using the random variable $Z$ defined by

\begin{eqnarray}
\label{eq:schemeP5}
	X(t) = \mu(t) + \sum_{k=1}^{\infty} z_k \xi_k (t),
\end{eqnarray}

\noindent where $\mu$ is the known average profile, the $\xi_k$ are orthonormal eigenfunctions that characterize variance of individual years from the mean, and the $z_k$ are uncorrelated random variables such that $E(z_k)=0$ and $V(z_k)=\lambda_k$ where $\lambda_k$ is the eigenvalues corresponding to the $k^{th}$ eigenfunction~\cite{Ramsay2005, Ramsay1991}. The eigenfunction pairs $\xi_k$, known as functional princpal components, are the leading eigenfunctions of the functional covariance defined by $v(s,t) = n^{-1}\sum_{i=1}^n (x_i(s)-\bar{x}(s))(x_i(t) - \bar{x}(t))$ and $z_k$ is an eigenvector of the Gram matrix $G$, defined by $G_{ij}= <x_i-\bar{x}, x_j - \bar{x}>$ with eigenvalues $n\lambda_k$.

\subsection{Interannual Canonical Correlation Analysis between LAI and Climate Attributes}

Consider a sample of paired annual curves $(x_1,y_1), \ldots, (x_m, y_m)$ from $(X, Y)$ where $X$ and $Y$ are random functions generating realizations of annual LAI and some climate attribute (either maximum temp or precipitation). The objective is to discover functions $(\xi, \eta)$ that maximize $cor(<x,\xi>, <Y, \eta>)$. Defining $Z=<\xi, X>$ and $W=<\eta, Y>$, gives

\begin{eqnarray}
\label{eq:schemeP6}
	\rho = cor(Z,W) = \frac{cov(Z,W)}{\sqrt{V(Z)V(W)}}
\end{eqnarray}

\noindent where 
\begin{align*}
 V(Z) =\int_a^b\int_a^b cov(X(s), X(t)\xi(s)\xi(t)dsdt = <\xi, cov(X(s),X(t))\xi> , \\ 
V(W) =\int_a^b\int_a^b cov(Y(s), Y(t)\eta(s)\eta(t)dsdt= <\eta, cov(Y(s),Y(t))\eta>  , \; \mbox{and}  \\
 cov(z,w) =\int_a^b\int_a^b cov(X(s), Y(t)\xi(s)\eta(t)dsdt = <\xi, cov(X(s),Y(t))\eta> 
\end{align*}

\noindent Subsequent canonical weight functions $(\xi_i, \eta_i)$ are found by maximizing Equation~\ref{eq:schemeP6} subject to 

\begin{align*}
cor(<x,\xi_j>, <x, \xi_i>)=0 , \\ 
cor(<y,\eta_j>, <y, \eta_i>)=0, \; \mbox{and}  \\
cor(<x,\xi_j>, <y, \eta_i>)=0 \\
\mbox{where} \; j=1,\ldots, i-1.
\end{align*}
 The functional canonical correlation analysis (fcca) requires some form of regularization to ensure meaningful weight functions $(\xi, \eta)$ since they only have $m$ constraints but as functions have infinite degrees of freedom (unlike classical CCA)~\cite{Ramsay2005}. This is accomplished by defining the smoothed sample curves using the first four functional principal components of $X$ and $Y$ as the basis expansion as opposed to a standard b-spline basis expansion. The fpca models have shown that the first 4 fpc's explain 90 to 99 percent of the variability of the original sample and are expressed by $x_i(t) = \mu_x + \sum_{k=1}^{4} s_{ik} \nu_k^{(1)}$ and $y_i(t) = \mu_y + \sum_{k=1}^{4} s_{ik} \nu_k^{(2)}$ where $\nu$ and $s_{ik}$ are defined similarly to Equation~\ref{eq:schemeP5}.

The strength of the relationship between LAI and temperature or precipitation, $\rho$, provides the conventional interpretation of correlation, but equally important, the paired weight functions $(\xi, \eta)$ are the components of variation that most account for the interaction of the two attributes.

%Materials and Methods should be described with sufficient details to allow others to replicate and build on published results. Please note that publication of your manuscript implicates that you must make all materials, data, computer code, and protocols associated with the publication available to readers. Please disclose at the submission stage any restrictions on the availability of materials or information. New methods and protocols should be described in detail while well-established methods can be briefly described and appropriately cited.

%Research manuscripts reporting large datasets that are deposited in a publicly avail-able database should specify where the data have been deposited and provide the relevant accession numbers. If the accession numbers have not yet been obtained at the time of submission, please state that they will be provided during review. They must be provided prior to publication.

%Interventionary studies involving animals or humans, and other studies require ethical approval must list the authority that provided approval and the corresponding ethical approval code.
%\begin{quote}
%This is an example of a quote.
%\end{quote}

%%%%%%%%%%%%%%%%%%%%%%%%%%%%%%%%%%%%%%%%%%
\section{Results}

\begin{figure}[H]	
%\widefigure
\centerline{\includegraphics[width=0.7\textwidth]{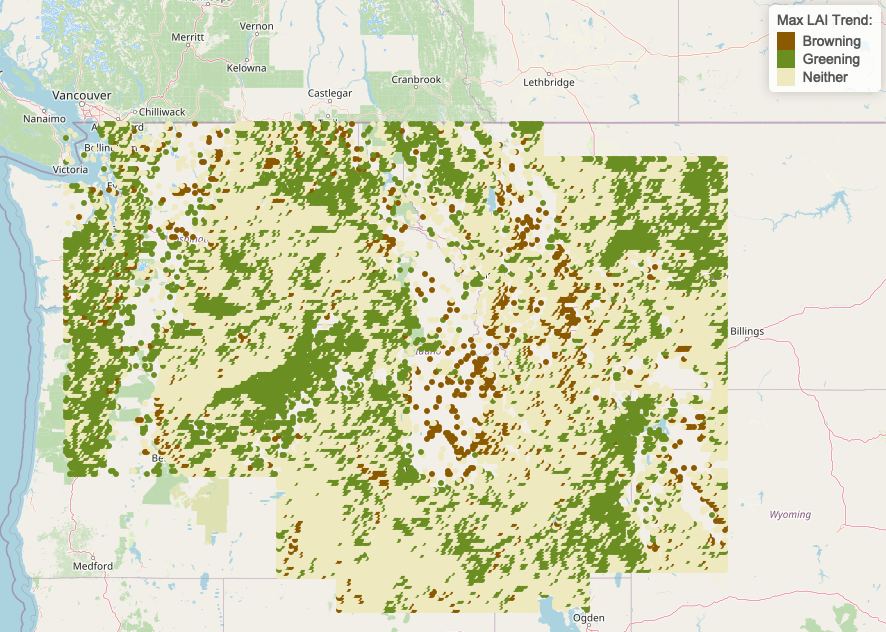}}
\caption{Temporal correlation of maximum annual LAI from 1996 to 2017 discretized to Greening, Browning, and Neither using a significance levels of $\alpha = 0.15$~\cite{Cheng2019}. Refer to the Supporting File S2 for applet access to examine trends across this region interactively.
\label{fig2}}
\end{figure}

This analysis detects widespread greening earlier in the growing seasons across a range of ecosystems in the CRB from 1996 to 2017. Initial exploration of correlations between annual maximum LAI and time in this region provide evidence of this phenological shift. Fig~\ref{fig2} and Tables ~\ref{tab1} and ~\ref{tab2}, show calculated correlations between detected annual maximum LAI at each site and time. Significant correlations were determined at $\alpha=0.15$. Sites that were significantly greening had positive correlations between LAI and time, whereas sites that were significantly browning had negative correlations between LAI and time. Non-significant correlations are denoted as \emph{neither} greening nor browning.
High frequencies of greening were detected along large sections of the the major CRB rivers (the Columbia and Snake) and in the Pacific coastal region of the CRB. Generous significant levels were chosen in this preliminary stage since, with only 22 replications of annual maximum LAI (over the 22 years), the correlation coefficients are sensitive to anomaly measurements and sub-intervals of decreasing annual maximum LAI. A high anomaly reading in early years makes it difficult to detect an increasing trend across the remaining years. Further, a sudden drop in maximum LAI in later years may leverage the correlation away from a high positive correlation detected in earlier intervals. 

The spatial distribution of greening magnitude and timing demanded a more rigorous exploratory analysis that accomplished the following objectives: (1) eliminate/filter anomalies (false high LAI recordings), (2) detect regions that are strongly correlated over time while retaining at least some information regarding spatial proximity, and (3) examine changes/perturbations in the functional structure of annual LAI profiles. 

\begin{figure}[H]	
%\widefigure
\centerline{\includegraphics[width=0.7\textwidth]{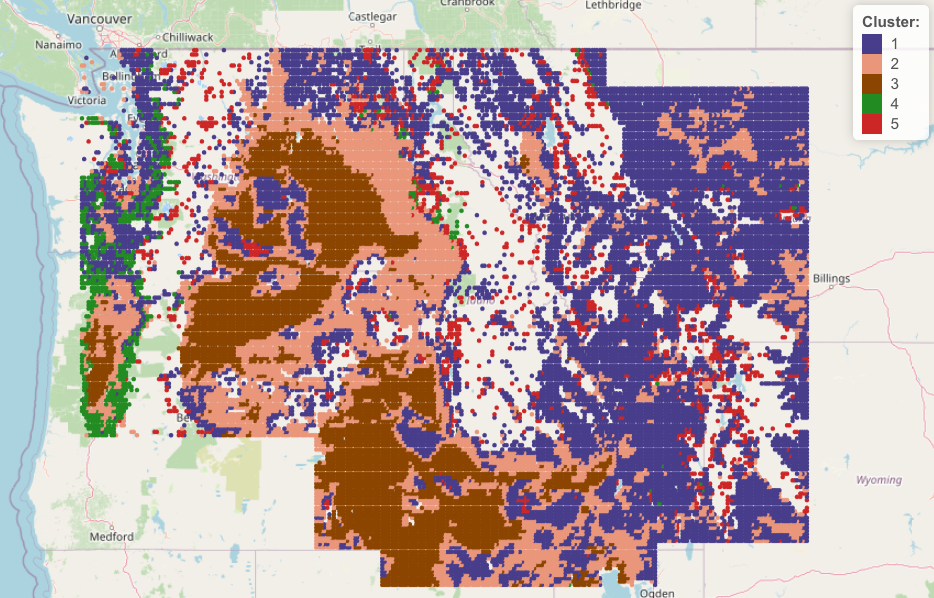}}
\caption{K-mediod cluster analysis of the pairwise correlation matrix of the 27191 B-spline smoothed LAI profiles~\cite{Cheng2019}. Refer to the Supporting File S2 for applet access to examine trends across this region interactively.
\label{fig3}}
\end{figure}

A k-mediods cluster analysis was performed on the 22-year B-spline smoothed LAI profiles using the dissimilarity matrix outlined in \ref{cluster}.  Sites allocated into the same cluster are determined to have strong multidecadal relationships to each other as inherited from the dissimilarity matrix. Fig~\ref{fig3} depicts the results of the 5 cluster k-mediods model geographically. The details of the all cluster models for $k=5$ are provided in Table~\ref{tab4}

The 5-clusters intuitively distinguish regions with different land cover, elevation, and climate characteristics based on satellite-derived LAI values. Tables ~\ref{tab1} and ~\ref{tab2}, show that across Clusters 1, 2, 4, and 5, 32 - 70 percent of sites are identified as greening sites, and extremely low percentages of sites in each cluster are identified as browning sites. Table~\ref{tab3}  shows the land cover classifications for each of the five clusters. Clusters 1,2, and 3 contain the largest proportion of agricultural sites and the majority land cover classification for these 3 clusters is Scrub/Shrub. Clusters 4 and 5 are dominantly forested evergreen sites. All clusters are distinguished by significant differences in elevation distributions, with Cluster 5 containing the highest elevation sites and Cluster 4 containing the lowest elevation sites(~\ref{figA1}).

\begin{figure}[H]	
%\widefigure
\centerline{\includegraphics[width=0.8\textwidth]{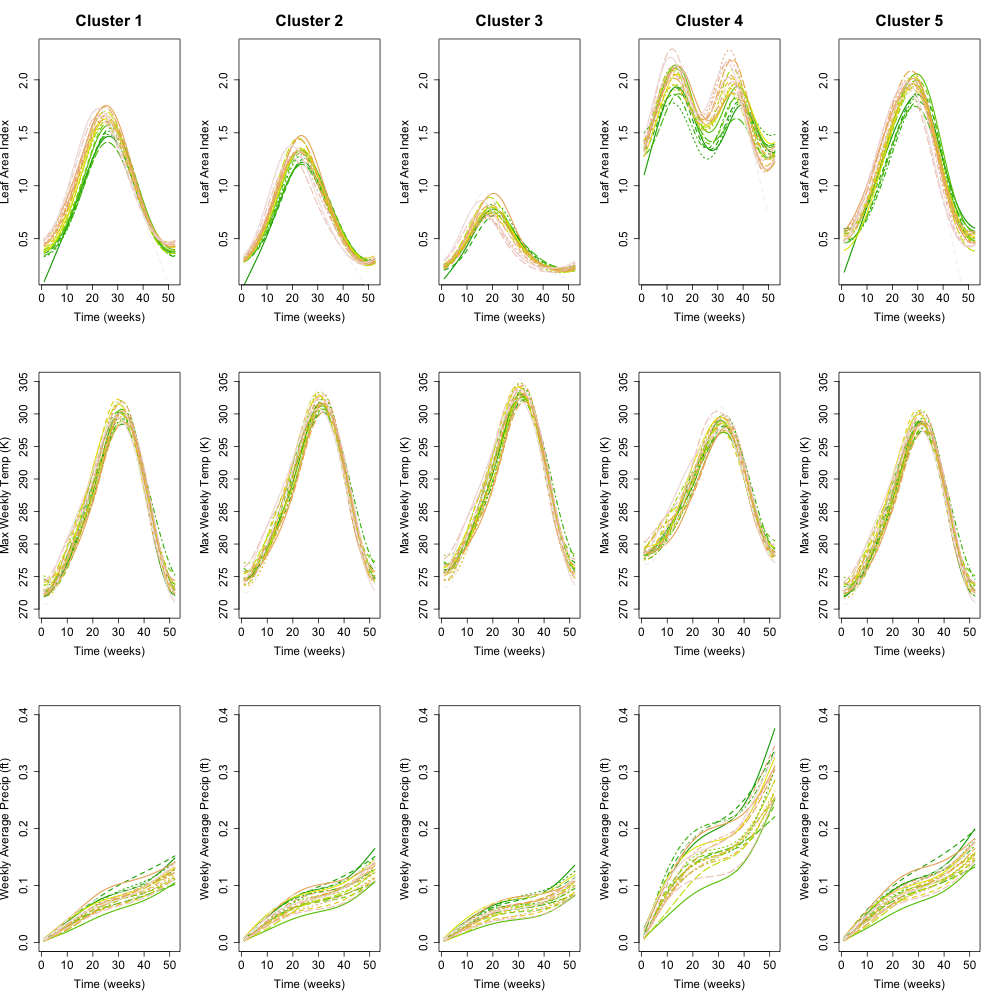}}
\caption{Interannual regional average weekly maximum LAI, maximum Temp, and average precipitation profiles. The curves are colored on a gradient scale where greener curves are closer to 1996 and pinker curves are closer to 2017. Noticeable time-dependent changes in LAI were identified where no clear trend in temperature and precipitation is visually observed.
\label{fig4}}
\end{figure}

Figure ~\ref{fig4} shows annual profiles for weekly maximum LAI, weekly maximum temperature, and average weekly precipitation for each of the five clusters. Annual maximum LAI is highest in the evergreen forested sites in Clusters 4 and 5 and lowest in Cluster 3, which contains sites with the highest proportion of Scrub/Shrub land cover in the CRB. Annual temperature profiles are similar throughout the CRB region, with slightly lower annual maximum and higher annual minimum temperatures detected throughout the 22-year time period at the low-elevation coastal sites comprising Cluster 4. Annual precipitation profiles are similar for Clusters 1,2,3, and 5, with Cluster 4 receiving much greater cumulative precipitation than the other clusters each year. 

\begin{figure}[H]	
%\widefigure
\centerline{\includegraphics[width=0.8\textwidth]{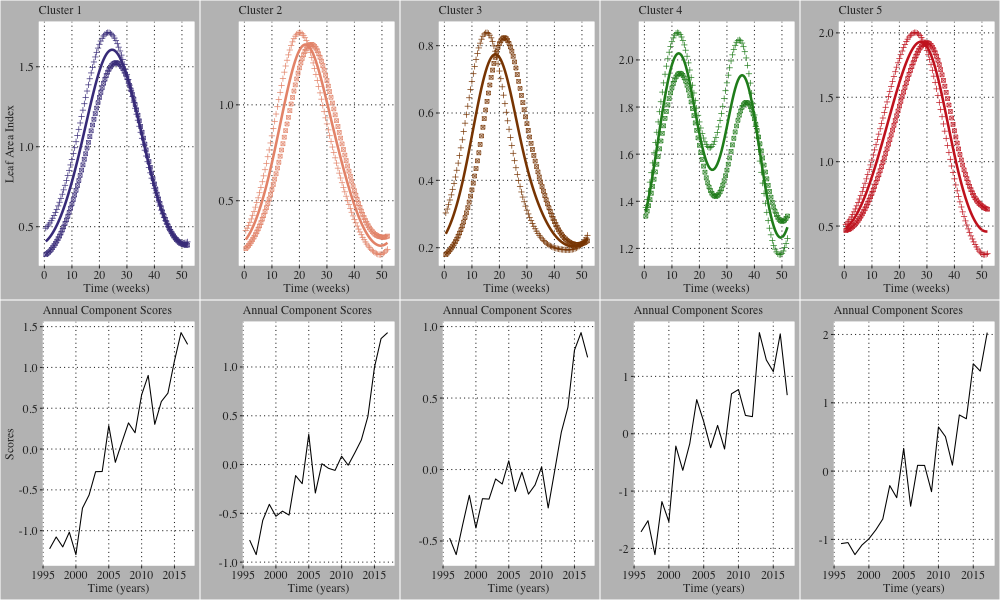}}
\caption{LAI functional principal components results for the first principal component of each cluster. Upper plots solid lines are the mean annual profiles and the (+) markers denote the trend line for the first component added to the mean function with an appropriate scaling factor, and the ($\square$) markers denote the trend line for the first component subtracted from the mean function with the same scaling factor. The lower plots visual the annual scores of the first component as a time series from 1996 to 2017. Years with scores greater than zero are characterized by the (+) trend line and years with scores less than zero are characterized by the ($\square$) trend line. 
\label{fig5}}
\end{figure}

\begin{figure}[H]	
%\widefigure
\centerline{\includegraphics[width=0.8\textwidth]{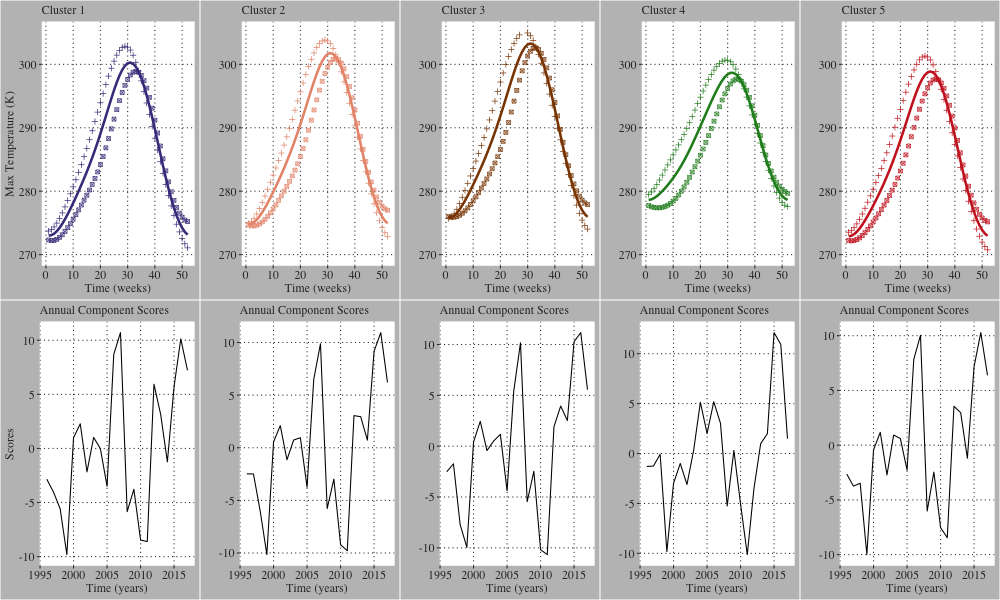}}
\caption{Weekly maximum temperature 1st functional principal component results for the first principal component of each cluster. 
\label{fig6}}
\end{figure}

\begin{figure}[H]	
%\widefigure
\centerline{\includegraphics[width=0.8\textwidth]{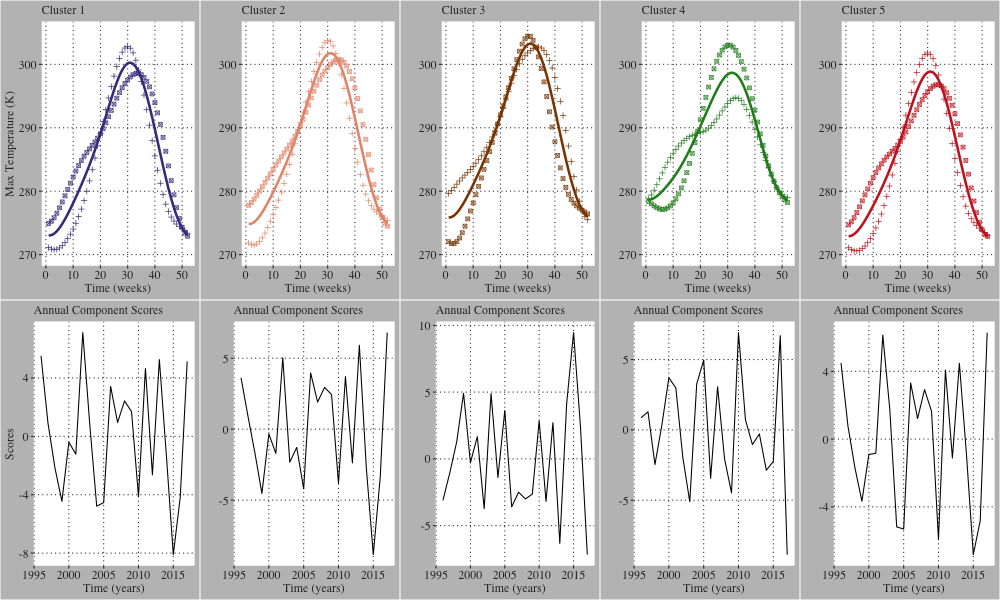}}
\caption{Weekly maximum temperature 2nd functional principal component results for the first principal component of each cluster. 
\label{fig7}}
\end{figure}

Functional principal components analysis on the annual LAI profiles found that among all of the clusters, 55-75 percent of the inter-annual variation is described by the first principal component characterized by an earlier and higher peak in annual maximum LAI. A linear increase in principal component scores over time indicates a trend toward earlier and higher annual maximum LAI throughout the CRB region  (Figure ~\ref{fig5}). This demonstrates that despite differences in land cover, elevation, and annual precipitation profiles between the five-clusters, a clear greening trend is detected over the 22-year period throughout the CRB. Identical analysis on annual cumulative precipitation showed ~90 percent of the inter-annual variation among all of the clusters was explained by the first principal component characterized by either greater or less annual cumulative precipitation. Functional principal components analysis on temperature showed ~45-48 percent of the inter-annual variability is explained by the first principal component associated with linearly warmer temperatures from the beginning of the year through the annual peak in summer temperatures. The second principal component, explaining ~20-23 percent of the inter-annual variability among the five clusters, is associated with either significantly warmer temperatures during roughly the first 20 weeks of the year and a lower annual maximum temperature, or cooler temperatures during the first 20 weeks of the year and a higher annual maximum. Annual maximum temperature and annual precipitation profiles do not demonstrate an obvious trend over the time period 1996-2017 (Figure ~\ref{fig6}, ~\ref{fig7}).

\begin{figure}[H]	
%\widefigure
\centerline{\includegraphics[width=0.8\textwidth]{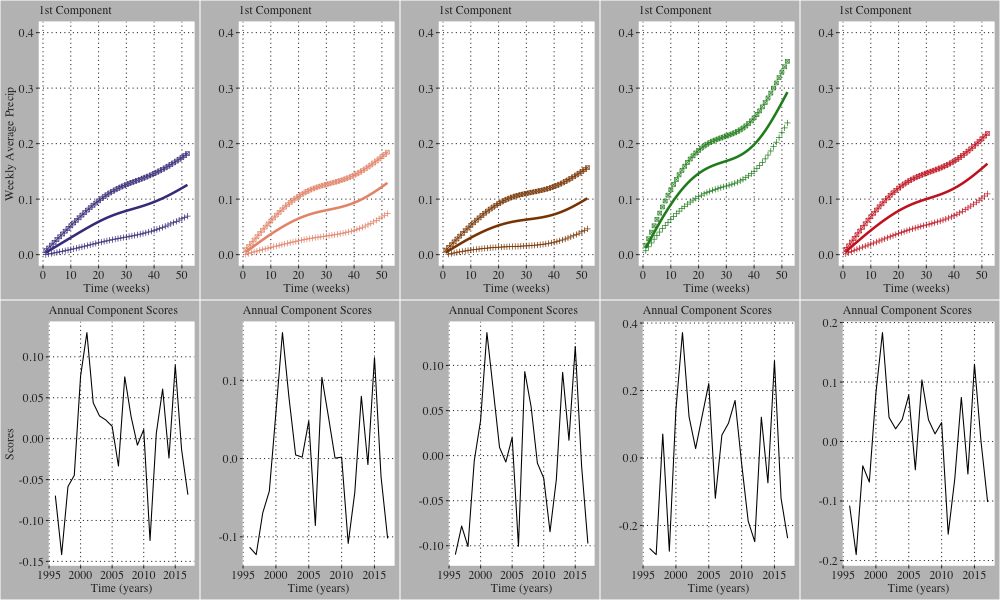}}
\caption{Weekly average precipitation first functional principal component results for the first principal component of each cluster. 
\label{fig8}}
\end{figure}

\begin{figure}[H]	
%\widefigure
\centerline{\includegraphics[width=0.8\textwidth]{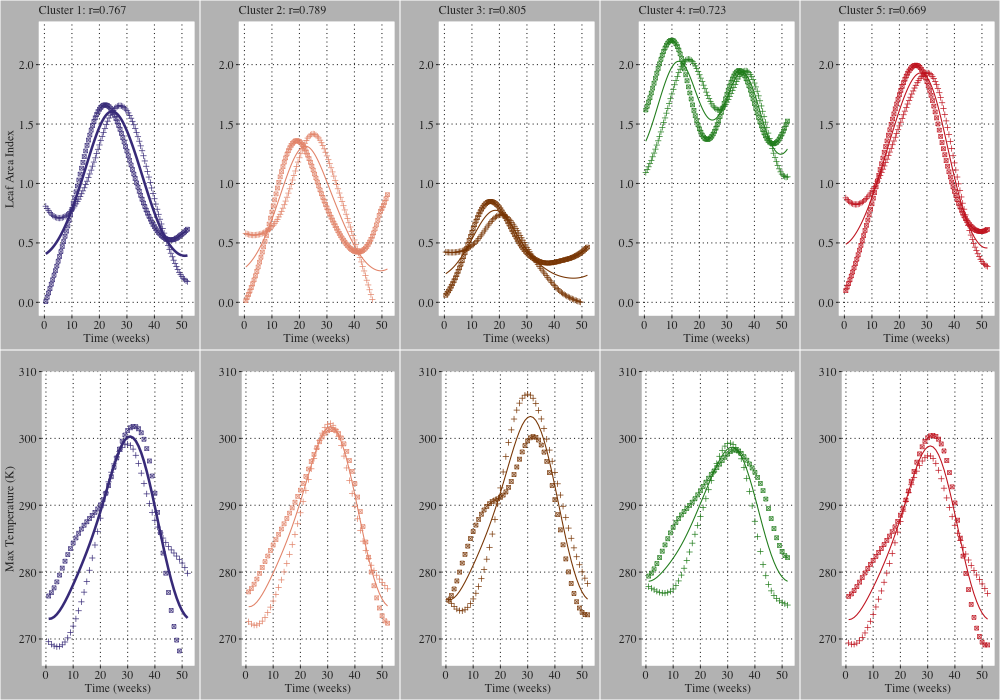}}
\caption{LAI vs Max. Temperature functional canonical correlation results for the first pair of canonical weight functions. The correlation between site attributes is listed above each columns of plots. The (+) markers denote the trend line for the first weight function (for either the LAI weight function or the maximum temperature weight function) added to the mean function with an appropriate scaling factor, and the ($\square$) markers denote the trend line for the first weight function subtracted from the mean function with the same scaling factor. The strength of the correlation between LAI and maximum temperature is characterized by examining the pair of (+)-profiles or ($\square$)-profiles across (for each site attribute). 
\label{fig9}}
\end{figure}

Functional canonical correlation analysis reveals correlations between intra-annual variation in temperature and precipitation and the earlier and higher LAI peak being detected in each of the clusters. In each cluster, the shift in phenology toward earlier and higher annual maximum LAI values are correlated with warmer temperatures during the first 20 weeks of the year, shown in Figure ~\ref{fig8}. Functional canonical correlation between LAI and precipitation did not yield a consistent correlation among the clusters. Greater and earlier maximum LAI was correlated with greater spring precipitation in Clusters 1 and 5, and greater but not earlier maximum LAI in Cluster 3 was also correlated with greater spring precipitation, shown in Figure ~\ref{fig9}. Taken together the results of the functional canonical correlations indicate that the widespread shift in phenology toward an earlier and higher peak in annual LAI in the CRB is largely associated with warmer temperatures early in the year. However, the differences in interannual trends in temperature and LAI over the studied time period suggest there are other drivers of the trend in LAI not captured in this analysis. Greater annual maximum temperatures are not shown to be correlated with this shift in LAI and the correlation between annual precipitation and the timing and height of the annual LAI peak varies between the 5 clusters.

%This section may be divided by subheadings. It should provide a concise and precise description of the experimental results, their interpretation as well as the experimental conclusions that can be drawn.

% The MDPI table float is called specialtable
\begin{table}[H] 
\caption{Summary of Max LAI temporal correlation results.\label{tab1}}
%%% \tablesize{} %% You can specify the fontsize here, e.g., \tablesize{\footnotesize}. If commented out \small will be used.
\begin{tabular}{ccccc}
%\toprule
\textbf{Regions}	& \textbf{Freq}	& Browning  &  Greening  & Neither   \\\hline
%\midrule
Cluster 1		& 12548		& 900  &   4087  &  7561 \\
Cluster 2		& 6737		& 104   &  2646 &   3987\\
Cluster 3		& 5593		& 82   &   790  &  4721 \\
Cluster 4		& 742		& 5   &   517   &  220 \\
Cluster 5		& 1571		& 207  &    547  &   817\\
%\bottomrule
\end{tabular}
\end{table}

% The MDPI table float is called specialtable
\begin{table}[H] 
\caption{Summary of the Max LAI annual location correlation results.\label{tab2}}
%%% \tablesize{} %% You can specify the fontsize here, e.g., \tablesize{\footnotesize}. If commented out \small will be used.
\begin{tabular}{cccc}
%\toprule
\textbf{Regions}	& Earlier  &  Later   & Neither  \\ \hline
%\midrule
Cluster 1		& 3885 &  157  &  8506 \\
Cluster 2		& 1398 &  119  &  5220\\
Cluster 3		& 1049  &  60  &  4484 \\
Cluster 4		& 82  & 102   &  558 \\
Cluster 5		& 349  &  26   & 1196\\
%\bottomrule
\end{tabular}
\end{table}

% The MDPI table float is called specialtable
\begin{table}[H] 
\caption{Summary Statistics of 5-cluster model. Proportion of land cover are zonal statistics from the MDA BaseVue 2013 Land Cover product with the coordinates of each cluster set as the zones. Elevation is extracted for each site from the USGS Ground Elevation digital terrain model.\label{tab3}}
%%% \tablesize{} %% You can specify the fontsize here, e.g., \tablesize{\footnotesize}. If commented out \small will be used.
\begin{tabular}{cccccc}
%\toprule
\textbf{Regions}	& \textbf{Freq}	& \textbf{Prop Agriculture}  & \textbf{Prop Scrub}   & \textbf{Prop Evergreen}   & \textbf{Med Elev (m)} \\ \hline
%\midrule
Cluster 1		& 12548		& 0.130      &   \textbf{0.293}    &   0.238    &  1451.1 \\
Cluster 2		& 6737		& 0.161      &   \textbf{0.455}    &   0.155    &  1150.3 \\
Cluster 3		& 5593		& 0.174      &  \textbf{0.667}    &   0.010    &  942.0 \\
Cluster 4		& 742		& 0.009      &   0.241    &   \textbf{0.481}    &  337.8 \\
Cluster 5		& 1571		& 0.022      &   0.209    &   \textbf{0.617}    &  1776.8 \\
%\bottomrule
\end{tabular}
\end{table}

\begin{table}[H] 
\caption{K-mediod 5-cluster model characteristics.\label{tab4}}
%%% \tablesize{} %% You can specify the fontsize here, e.g., \tablesize{\footnotesize}. If commented out \small will be used.
\begin{tabular}{cccccc}
%\toprule
\textbf{Regions} & \textbf{Freq}	& \textbf{Max Diss}	& \textbf{Avg Diss}  & \textbf{Diameter}   & \textbf{Separation}  \\\hline
%\midrule
Cluster 1		& 12548		& 0.6775 & 0.0744 & 1.0174  &  0.0025 \\
Cluster 2		& 6737		& 0.6746 & 0.0730 & 1.0496  &  0.0025 \\
Cluster 3		& 5593		& 0.8158 & 0.0956 & 1.1354  &  0.0033 \\
Cluster 4		& 742		& 0.9487 & 0.3087 & 1.2693  &  0.1110 \\
Cluster 5		& 1571		& 0.8639 & 0.2350 & 1.1874  &  0.0035 \\
%\bottomrule
\end{tabular}
\end{table}

\begin{table}[H] 
\caption{Percent of variation explained by the first principal component of LAI, Maximum Temperature, and precipitation.\label{tab5}}
%%% \tablesize{} %% You can specify the fontsize here, e.g., \tablesize{\footnotesize}. If commented out \small will be used.
\begin{tabular}{cccc}
%\toprule
\textbf{Regions} & \textbf{Attribute}   & \textbf{Proportion of Variation}	 & \textbf{Component}\\\hline
%\midrule
Cluster 1		&  LAI  & 0.757 & 1st\\
Cluster 2		&  LAI  & 0.639 & 1st\\
Cluster 3		&  LAI  & 0.554 & 1st\\
Cluster 4		&  LAI  & 0.647 & 1st\\
Cluster 5		&  LAI  & 0.604 & 1st\\
Cluster 1		&  Max Temp  & 0.475 & 1st\\
Cluster 1		&  Max Temp  & 0.214 & 2nd\\
Cluster 2		&  Max Temp  & 0.481 & 1st\\
Cluster 2		&  Max Temp  & 0.210 & 2nd\\
Cluster 3		&  Max Temp  & 0.482 & 1st\\
Cluster 3		&  Max Temp  & 0.198 & 2nd\\
Cluster 4		&  Max Temp  & 0.446 & 1st\\
Cluster 4		&  Max Temp  & 0.233 & 2nd\\
Cluster 5		&  Max Temp  & 0.462 & 1st \\
Cluster 5		&  Max Temp  & 0.225 & 2nd\\
Cluster 1		&  Precip  & 0.903 & 1st\\
Cluster 2		&  Precip  & 0.914 & 1st\\
Cluster 3		&  Precip  & 0.917 & 1st\\
Cluster 4		&  Precip  & 0.917 & 1st\\
Cluster 5		&  Precip  & 0.912 & 1st\\
%Cluster 1		&  Precip  & 0.384\\
%Cluster 2		&  Precip  & 0.416\\
%Cluster 3		&  Precip  & 0.397\\
%Cluster 4		&  Precip  & 0.438\\
%Cluster 5		&  Precip  & 0.433\\
%\bottomrule
\end{tabular}
\end{table}

% NOTE add percent variation explained to these plots

\begin{figure}[H]	
%\widefigure
\centerline{\includegraphics[width=0.8\textwidth]{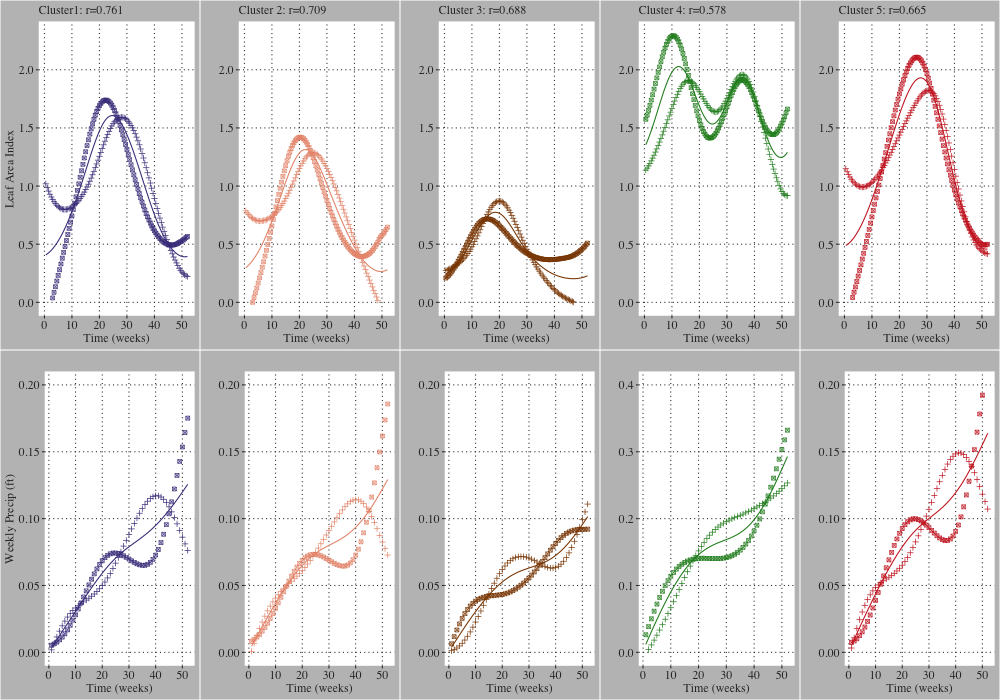}}
\caption{LAI vs Cumulative Precipitation functional canonical correlation results for the first pair of canonical weight functions. The correlation between site attributes is listed above each columns of plots. The (+) markers denote the trend line for the first weight function (for either the LAI weight function or the precipitation weight function) added to the mean function with an appropriate scaling factor, and the ($\square$) markers denote the trend line for the first weight function subtracted from the mean function with the same scaling factor. The strength of the correlation between LAI and precipitation is characterized by examining the pair of (+)-profiles or ($\square$)-profiles across (for each site attribute). 
\label{fig10}}
\end{figure} 

\section{Discussion}

% Power of using methods we've used
    The results of this work demonstrate the utility of FDA for the detection of annual greening trends of high dimensional phenological processes, and more specifically, the characterization of the within-year and across-year trends in vegetation dynamics. Annual greening of field measured or remotely sensed sites is best characterized by multiple parameters, namely the magnitude of the peak "greenness", the (annual) timing of the peak, the duration of greeness, and the point of maximum change in greenness. Although only the first and second of these parameters are examined in this analysis, all of these features are inherently contained in the sample of 27196 smooth functions used in this study. Any future work to examine the other features of LAI curves can be performed using the same preprocessing used here. Without the use of spline smoothed LAI curves, the analysis of these individual parameters must be assessed without the same theoretical cohesiveness present in this approach. Ultimately, annual LAI profiles are simple to smooth, and we demonstrate that variation in annual LAI across years is effectively detected and explained using functional principal components analysis.

We believe that the modeling of processes with underlying continuity should take advantage of this continuity when possible. In this analysis, we demonstrate that the modeling of continuous processes in the presence of high volumes of missing values is achievable, and our results lead us to believe that our choice to only consider sites with less than 72 percent missing values is conservative, and such an analysis would be effective with upwards of 80 percent missing values. This provides opportunities to use such methods in regions where remotely sense greenness indices are recorded in with extreme sparseness, such as boreal, and Arctic climates. We must acknowledges disadvantages of our sampling of sites in the region. First, the quality of the LAI AVHRR CDR has improved in quality across its entire domain from 1982 to present. Beginning in 1996 eliminates some of the years with the lowest quality, but it is possible and likely that we have removed sites that had improved satellite coverage in the later years of the domain. Also, we note that the removal of most sites along the Montana-Idaho border and mountains of Oregon and Washington is not indicating low-greenness but rather poor coverage through a substantial portion of years in our study.

Our clustering approach used in this project is effective at separating regions intuitively across an array of variables (land cover, elevation, temperature, and precipitation) with only the use of satellite-derived LAI profiles, and further, this provides the theoretical framework used to make regional inferences about changes in climate and greenness. The approach is proficient for analyzing tens or hundreds of thousands of sampled sites. 

Although our cluster model retains some implicit use of proximity of sites (since closer sites have a tendency to have higher correlations), we believe that there are necessary improvements to such a clustering approach, namely the filtering of noise sites (using methods such as DBSCAN) and inducing a spatially weighting (or penalization) on the dissimilarity matrix used in our work. We argue that the merits of the approach taken justify its presentation here, and we encourage further work in unsupervised learning methodology of phenological processes.

We emphasize that our work here is strictly exploratory, and not predictive. The relationships between climate and LAI discussed here are associative by nature, and further work using functional regression models is required to explore the fascinating predictive relationship between these attributes. Recent literature implementing predictive modeling of fpca scores of NDVI has yielded promising results, and this approach can be extended to model climatic factors (such as CO2 concentration, GPP, plant respiration, Soil moisture) that predict higher fpca scores for LAI in recent years in this region~\cite{Pesaresi2020}.

Our work is also insufficient without recognizing disadvantages of the LAI CDR using AVHRR sensors~\cite{Fensholt2002, Steven2003,Hansen2002} . Recent literature has shown that this product is lesser to MODIS sensors in making valid inferences on field measurement resolution and areas with higher annual precipiation ($>$1m) precipitation~\cite{Fensholt2009,Cihlar2001}. As shown in Fig~\ref{fig4}, regional cumulative precipiation averages for all clusters are well below this threshold. We add our work to the body of literature on the detection of changes in remotely sensed greening, and we emphasize that the methods used in this project are directly extendable to any remotely sensed time-series data.

% Functional Data methods advantages (namely we see clear greening!!) capturing interannual variability.

% Plant responses to climate change
  In this analysis, we were able to reveal a shift in annual vegetation dynamics from 1996-2017 across a range of land cover classes and ecosystem types in the CRB of North America. Plant phenological events in temperate regions are triggered predominantly by the well-known climatic changes associated with the changing of the seasons. These responses of vegetation to environmental conditions provide a measurable and accurate signature of the impacts climate change is having on plants ~\cite{Parmesan and Yohe 2003}. The importance of understanding how plants respond to changing climate conditions has led to considerable work on the influence of climate variables on plant phenology ~\cite{Fitchett 2015}. Our analysis investigates the intra-annual relationships between vegetation dynamics and the climate variables temperature and precipitation throughout the year over a multidecadal timescale. Temperature is the dominant driver of the timing of many plant developmental processes and phenological shifts ~\cite{Piao 2015, He 2015, Menzel 2006}. Plants synchronize their growth and development with favorable thermal conditions in order maximize the growing season and minimize the risk of frost damage. Sufficient exposure to cold temperatures in the winter is required for many plants to break dormancy, and a subsequent accumulation of degree days in the spring (time above a given temperature threshold) triggers budburst and the unfolding of leaves ~\cite{Harrington 2010}. 

In the present study, earlier and higher annual maximum LAI throughout the CRB was largely correlated with higher temperatures during the first 20 weeks of the year. Phenological responses to environmental conditions can vary significantly among different regions and plant species ~\cite{Sherry 2007, Richardson 2013}; however, similar trends in vegetation dynamics were found across a range of natural ecosystem and land cover types in the CRB, indicating common responses to abiotic environmental factors across the regional scale of this study. On agricultural land, spring planting date could influence the timing and magnitude of the peak in annual LAI. However, each of the 5 clusters contains a variety of land cover types and relatively small proportions of agricultural land. Because all of the clusters are showing coherent trends in LAI over the time period studied, the effect of differences in planting date on the agricultural fields in each cluster likely does not significantly influence the satellite-observed regional vegetation dynamics in the CRB.

The same intra-annual temperature trend was correlated with the earlier and higher maximum LAI values in each of the five clusters in the CRB. This result shows that greater accumulation of warm temperatures early in the year leads to an earlier onset of budburst and leaf unfolding, as well as an earlier peak of plant productivity in the summer growing season.  Intra-annual trends in cumulative precipitation did not demonstrate a uniform correlation with the observed LAI trend among the five clusters in the CRB. This aligns with  previous research that shows differential responses of phenology to precipitation between arid and wet regions and an overall lesser or indirect contribution of precipitation to plant phenological shifts compared to temperature ~\cite{Shen 2015, Morin 2010}. 

A global warming trend of ~0.2 degrees C per decade has been observed since the 1980's ~\cite{Hansen 2006}, and recent warming of the Northern Hemisphere, particularly in the winter and spring, is well documented ~\cite{Schwartz 2006, Robeson 2004}. Results of the functional principal components analysis on temperature in the CRB showed that greater than 60 percent of the interannual variability can be explained by warmer temperatures early in the year. Despite this, significant interannual variations in temperature trends exist across the CRB region between 1996 and 2017. Although warmer temperatures early in the year are seemingly the most important factor influencing the greening trend over time in this analysis, a clear trend toward early-year warming over the time period studied is lacking as shown by the lack of linearly increasing principal component scores for temperature over the 22 years. This indicates that while warmer spring temperatures are clearly influential over vegetation dynamics in the CRB, there are likely other factors playing important roles in the observed greening trend over time.

Vegetation dynamics in most plant species are mainly governed by temperature, photoperiod, precipitation, and the interactions among these key variables. The sensetivity of phenological shifts to these climate variables can differ among regions and plant species (especially sensitivities to photoperiod and precipitation ~\cite{Adole 2019, Ghelardini 2010, Morin 2010}), and many of the underlying biological mechanisms that control phenological responses to these climate variables are still unknown. The influence of different climate variables have on plant phenology are entangled and the combined effects likely promote or constrain observed trends in vegetation dynamics ~\cite{Jolly 2005, Garonna 2018}. For example, in mid and high latitude regions, warming temperatures in the spring are correlated with increasing day length. In photoperiod-sensitive plant species, early warming before the a particular daylength threshold is reached could constrain the temperature effect on spring phenology. Also, clouds associated with heavy spring precipitation could lower the sunlight intensity and quality and similarly constrain spring phenology. The timing of snowmelt is also an important factor in spring phenological shifts in regions with cold winters ~\cite{Shutova 2006, Lambert 2010}. Timing of snowmelt is a function of both temperature and winter precipitation and the depth and persistence of snow cover can effect the ability of plants to respond to changing photoperiod and air temperature early in the year. In parts of the CRB where snowcover persists through the winter, earlier snowmelt due to either warmer spring temperatures, less winter precipitation, or both, could also be correlated with regional greening trends. Further investigation of the interactive and combined effects of climate variables on vegetation dynamics in the CRB is needed to fully understand the relationship between changing environmental conditions and observed trends in LAI. 

The effect of globally increasing concentrations of atmospheric CO\textsubscript{2} on vegetative growth can also not be overlooked. Increases in anthropogenic emissions and land use change since the industrial revolution has driven the atmospheric CO\textsubscript{2} concentration to over 400 parts per million (ppm), a roughly 40 percent increase since pre-industrial times ~\cite{IPCC 2013}. Higher concentrations of CO\textsubscript{2} in the atmosphere suppress the oxygenase activity of the main carbon-fixing enzyme in plants, Ribulose 1,5-bisphosphate carboxylase-oxygenase (Rubisco). This leads to reduced rates of the carbon and energy dissipative process of photorespiration and increased photosynthetic carbon assimilation. While increasing concentrations of atmospheric CO\textsubscript{2} are not found to change the timing of annual plant phenology ~\cite{H"anninen 2007}, greening trends around the world have been attributed to the "fertilization effect" of increasing concentrations of atmospheric CO\textsubscript{2} ~\cite{Zhu 2016, Donohue 2013, Haverd 2020}. Though not investigated in the present study, increasing atmospheric [CO\textsubscript{2}] could play a role in the increasing magnitude of annual maximum LAI observed in the CRB.

%ecosystem services and function: including changes in carbon sequestration and water balance. 
%The alteration in the timing of the growing season can effect the ranges of different plant species and have an impact on plant-animal interactions in the area. 
%Agricultural production in the CRB is also likely effected by the greening trend. 
%affect human activities- recreational suitability of landscape, pollen flight forecasts, etc

%The present study provides a useful framework for investigating vegetation responses to changing climate conditions using remotely sensed data

% To date/ to the best of our knowledge there are not much literature on interannual variability that investigates variation continuously through the year.
% LAI is simple process, changes in the shape of the curve are basic and easy to model using these techniques and we can capture a lot of variation because of that.

%%%%%%%%%%%%%%%%%%%%%%%%%%%%%%%%%%%%%%%%%%
\section{Conclusions}

Our analysis detects a trend toward earlier and higher annual maximum LAI across a large portion of the Northwestern United States and strong associations with important climate variables between 1996 and 2017. We  detect this trend and these associations holistically using all available weekly measurements of LAI at each site to derive smooth LAI curves which retain critical annual attributes. This greening trend is detected across a variety of natural ecosystems and land cover types. Shifting vegetation dynamics in the CRB could have implications on ecosystem functions and services, plant-animal interactions and distributions, agricultural production, and human activities in the area, and we encourage further work to explore best practices in the analysis of vegetation dynamics to promote improvements to environmental and agricultural policy development.

%%%%%%%%%%%%%%%%%%%%%%%%%%%%%%%%%%%%%%%%%%
\vspace{6pt} 

%%%%%%%%%%%%%%%%%%%%%%%%%%%%%%%%%%%%%%%%%%
%% optional
%\supplementary{The following are available online at \linksupplementary{s1}, Figure S1: title, Table S1: title, Video S1: title.}


%%%%%%%%%%%%%%%%%%%%%%%%%%%%%%%%%%%%%%%%%%

\section*{Supporting Information}

%These commands reset the figure counter and add "S" to the figure caption (e.g. "Figure S1"). This is in case you want to add actual figures and not just captions.
\setcounter{figure}{0}
\renewcommand{\thefigure}{S\arabic{figure}}

% You can use the \nameref{label} command to cite supporting items in the text.
\subsection*{S1 Figure}
\begin{figure}[H]	
%\widefigure
\centerline{\includegraphics[width=0.7\textwidth]{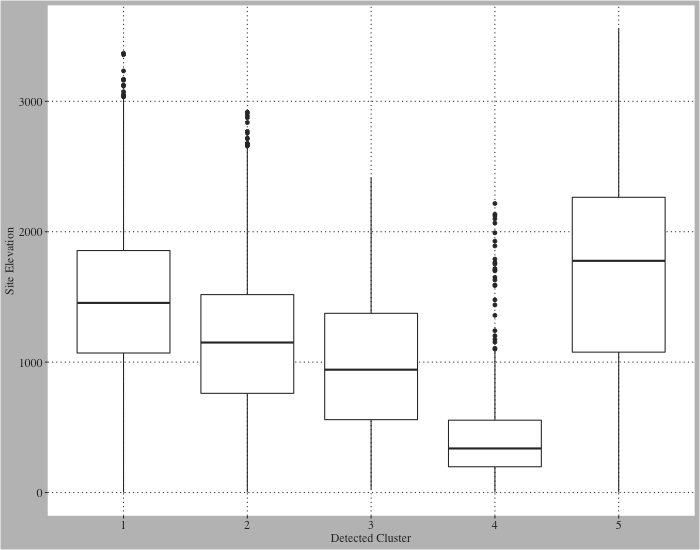}}
\caption{Distribution of site elevations by Cluster. We conducted one-way anova testing and post-hoc Tukey adjusted comparison testing of cluster means, and we detected highly significant ($<$0.0001) differences in the distribution of elevation across clusters.  
\label{figA1}}
\end{figure} 

\subsection*{S2 Applet}
\label{S2Applet}
{\bf CRB LAI Exploration Applet.} Click \href{https://abwhetten.shinyapps.io/CRB_LAI_1996_2017/}{abwhetten.LAI.CDR} to access applet.

%\clearpage

\section*{Acknowledgments}
We would like to thank members of the USDA ARS Lisa Ainsworth Lab for their continuing discussion and feedback on our project.

\end{document}